\documentclass[aps,twocolumn,prl,tightenlines]{revtex4}
\usepackage[dvips]{graphicx}
\usepackage[english]{babel}
\usepackage{amsmath}
\usepackage{amssymb}

\begin{document}

\title{Measuring charge based quantum bits by a superconducting single-electron transistor}
\author{J.Kinnunen, P.T\"orm\"a and J.P.Pekola}
\affiliation{Department of physics, P.O.Box 35 (YFL), FIN-40014, University of Jyv\"askyl\"a, Finland}
\begin{abstract}
Single electron transistors have been proposed to be used as a read-out device for Cooper pair charge qubits. Here we show
that a coupled superconducting transistor at a threshold voltage is much more effective in measuring the state of a qubit than a normal-metal 
transistor at the same voltage range. The effect of the superconducting gap is to completely block the current through the transistor
when the qubit is in the logical state 1, compared to the mere diminishment of the current in the normal-metal case. The time evolution of the 
system is solved when the measuring device is driven out of equilibrium and the setting is analysed numerically for parameters 
accessible by lithographic aluminium structures.
\end{abstract}

\maketitle

Nanoscale devices such as Cooper pair boxes or coupled quantum dots have been suggested as scalable and integrable realisations of quantum bits.
The two logical states of a qubit are the different charge states, or in the case of a flux qubit, flux states of the system. 
There are several proposals for quantum gates \cite{Schoen2000, Averin1998} and for interqubit couplings \cite{Makhlin1999},
as well as for measuring devices \cite{Bouchiat1998, Nakamura1999, Aassime2001, Vion2002}.
Permanently coupled \emph{normal-metal} transistors have been suggested as a device for measuring the state of a Cooper pair charge qubit \cite{Schoen1998, Averin2000}. 
Also, a \emph{superconducting} SET in the Josephson current regime (low biasing voltage) has been experimentally tested \cite{Cottet}. 
In this work we show that in the regime of high biasing voltages, the superconducting SET \cite{Korotkov1996, Averin1997} leads 
to a highly efficient quantum nondemolition measurement \cite{Khalili1995} due to the blocking effect of the gap.

\begin{figure}
        \includegraphics[scale = 0.80]{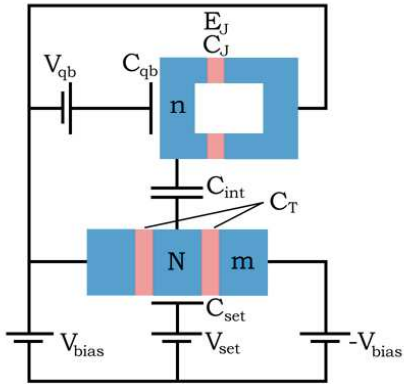}
        \caption{A Cooper pair charge qubit is capacitively coupled to a measuring single-electron transistor. The quantum numbers $n$, $N$ and $m$ are explained in the text.
	     The SET is symmetrically biased with voltage $V_{\mathrm{bias}}$, and the voltage $V_{\mathrm{set}}$ is the gate voltage.
                   The energy scales are determined by the interaction capacitance $C_ {\mathrm{int}}$, the tunnel junctions' capacitances $C_\mathrm{T}$ and the gate capacitance $C_{\mathrm{set}}$.}
        \label{fig:Setting}
\end{figure}

The setting is shown in Fig.~\ref{fig:Setting}. In the upper part, the Cooper pair box forms the qubit, its state characterised by the number of excess Cooper pairs
in the box, $n$. In the lower part, the superconducting single-electron transistor is capacitively coupled to the qubit with its state characterised by the excess charge
on the island, $eN$. In addition, the quantum number $m$ counts the number of charges passing through the SET in left-to-right direction. Without a biasing voltage 
$V_{\mathrm{bias}}$ across the SET, there is no dissipative current and no information is received. Moreover, in the absence of a dissipative environment, no dephasing 
of the composite system will occur and quantum operations on the qubit can be performed.

In order to perform the measurement, a biasing voltage is applied. As different qubit eigenstates correspond to different conductance in the SET, by observing the current 
one receives information on the state of the qubit. The time needed for the current to give the essential information 
is called the \emph{measurement time}. The back action caused by the SET dephases the qubit and eventually destroys also the logical state of the qubit.
The corresponding time scales are called the \emph{dephasing} and the \emph{mixing time}, respectively. For a good non-demolition measurement of the logical state ($|a|^2$ and $|b|^2$ in $a |0\rangle + b |1\rangle$), 
one expects to have a much longer mixing than a measurement time scale. We show that for a superconducting SET, in a parameter range accessible by aluminium structures, the ratio between the mixing and 
measurement time scales is in excess of 600.

The total Hamiltonian consists of three parts: the Hamiltonians of the SET, the qubit and the interaction
$H_\mathrm{set}$, $H_\mathrm{qb}$ and $H_\mathrm{int}$, respectively. The SET Hamiltonian is defined as
\begin{equation}
\label{eq:hamset}
	H_\mathrm{set} = E_\mathrm{set} \left (N - Q_\mathrm{set} \right)^2 + H_\mathrm{T} + H_\mathrm{L} + H_\mathrm{R} + H_\mathrm{I},
\end{equation}
where $E_\mathrm{set}$ is the charging energy, $Q_\mathrm{set}$ the gate charge of the transistor and 
\begin{equation}
        	H_\mathrm{T} = \sum_{kk' \sigma} T_{kk' \sigma}^{\mathrm{L}} c_ {k \sigma}^{\dagger \mathrm{I}} c_{k' \sigma}^{\mathrm{L}} e^{i \phi} + T_{kk' \sigma}^{\mathrm{R}} c_ {k \sigma}^{\dagger \mathrm{R}} c_{k' \sigma}^{\mathrm{I}} e^{-i \phi}e^{i \Psi} + H.c.
\end{equation}
describes quasiparticle tunnelling within the SET. The phase differences $\phi$ and $\Psi$ are the conjugate variables of $N$ and $m$, respectively. Thus, the operator
$e^{i \phi}$ ($e^{i \Psi}$) corresponds to quasiparticle tunnelling onto the island (right electrode) of the SET, increasing the quantum number $N$ ($m$) by one.
The last three terms in Eq.~(\ref{eq:hamset}) are defined as
\(
        H_r = \sum_{k \sigma} \epsilon_{k \sigma} c_{k \sigma}^{\dagger r} c_{k \sigma}^r, \text{   $(r = \mathrm{L,R,I})$},
\)
where $\sigma$ labels the transverse channels including the spin and $k$ labels the wave vector within one channel. These describe the noninteracting electrons in the 
left electrode, the right electrode and the island, respectively. Cooper pair tunnelling is excluded from the SET Hamiltonian due to the high biasing voltage. However,
this approximation is not valid for a low voltage, and the contribution from the Cooper pair tunnelling should be added when analysing the effect of the SET on the qubit during
logical operations.

The qubit Hamiltonian is defined as
\begin{equation}
\label{eq:hamqb}
	H_{\mathrm{qb}} = E_{\mathrm{qb}} \left( n - Q_{\mathrm{qb}} \right)^2 - E_\mathrm{J} \cos \Theta,
\end{equation}
where $E_\mathrm{qb}$ is the charging energy, $Q_\mathrm{qb}$ qubit's gate charge and the last term describes the transfer
of Cooper pairs to and from the box ($e^{i\Theta} |n\rangle = |n+1\rangle$). The quasiparticle tunnelling in the qubit is suppressed by the
Coulomb blockade, and the microscopic degrees of freedom have already been integrated out.
The interaction Hamiltonian describes the Coulomb interaction between the qubit and the transistor and is defined as
\begin{equation}
        H_\mathrm{int} = E_\mathrm{int} nN,
\end{equation}
where $E_\mathrm{int}$ is the charging energy. All charging energies $E_\mathrm{int}$, $E_\mathrm{set}$ and $E_\mathrm{qb}$ are 
determined by the capacitances of the system, and the gate charges $Q_\mathrm{qb} = - eV_n/E_\mathrm{qb}$ and $Q_\mathrm{set} = -eV_N/2E_\mathrm{set}$ 
depend on the effective gate voltages $V_n$ and $V_N$ that, for a symmetric bias, are determined by the gate voltages $V_\mathrm{qb}$ and $V_\mathrm{set}$ 
as given in \cite{Makhlin1999}.

Concentrating on the values of $Q_\mathrm{qb}$ around the degeneracy point $Q_\mathrm{qb} = \mbox{$\frac{1}{2}$}$, only the low-energy
charge states $n=0$ and $n=1$ are relevant. This assumption is required for the Cooper pair box to constitute a quantum bit. In the qubit's two-state approximation, 
the diagonalised operator becomes
$H_{\mathrm{qb}} = -\text{\mbox{$\frac{1}{2}$}} \Delta E \sigma_z$, where
\begin{equation}
\label{eq:Qqb}
        \Delta E =  \sqrt{\left[ E_{\mathrm{qb}}(1-2Q_{\mathrm{qb}}) \right]^2+E_J^2},
\end{equation}
while the operator $n$ becomes nondiagonal \(n = \mbox{$\frac{1}{2}$} \left( 1 - \cos (\eta) \sigma_z - \sin (\eta) \sigma_x \right) \),
where the mixing angle is given by $\tan \eta = E_\mathrm{J}/\left[ E_{\mathrm{qb}}(1-2Q_{\mathrm{qb}}) \right]$. Assuming the gate 
charge $Q_{\mathrm{qb}}$ sufficiently different from the degeneracy point $0.5$, the mixing angle $\eta$ is small in the charge-qubit 
approximation $E_\mathrm{J} \ll E_{\mathrm{qb}}$.
By rearranging the operators, the final form for the total Hamiltonian can be written as \( H = H_0 + H_\mathrm{T}\), where
\begin{equation}
\label{eq:Qset}
        H_0 = H_\mathrm{L} + H_\mathrm{R} + H_\mathrm{I} + H_{\mathrm{qb}} + E_{\mathrm{set}} \left( N-Q_{\mathrm{set}} \right)^2 + H_\mathrm{int}.
\end{equation}

We analyse the measurement process by master equation techniques. The master equation for the system reads
\begin{equation}
\label{eq:master}
        \frac{ \partial \sigma (t) }{ \partial t } + \frac{i}{ \hslash }\left[ H_0, \sigma (t) \right] = \mathrm{Tr}_\mathrm{L,R,I} \int_0^t \Sigma (t-t') \sigma (t') dt',
\end{equation}
where the trace is taken over the microscopic degrees of freedom of the transistor's left and right electrodes and the island. The elements of the transition matrix
\( \Sigma (t-t') = - \frac{1}{ \hslash^2 } \left( \left[V, U(t-t') \left[V, \cdot \right] U(t'-t) \right] \right) \)
are calculated by using the diagrammatic technique developed in \cite{Schoen1994} and \cite{Makhlin1999}. 
However, the rules for converting diagrams into integral
equations must be changed to include the superconducting density of states
\begin{equation}
\label{eq:dens}
       N(x) =  \begin{cases}
                        \frac{|x|}{\sqrt{ x^2-{\Delta}^2}}N_0, & \text{if $|x| > \Delta$} \\
                        0, & \text{otherwise,}
                   \end{cases}
\end{equation}
where $\Delta$ is the superconducting gap. In zero-temperature limit, the matrix elements $\Sigma (t-t')$ can be written as
\begin{equation}
	\Sigma_{k,l} (t-t') = F(t-t')^2 e^{-iE(t-t')/\hslash},
\end{equation}
where \( F(t) = \int_{0}^{\infty} N(x) e^{-(\varepsilon+it)x} dx \),
and $\varepsilon$ is a cut-off exponent which ensures the convergence of the integral. With the superconducting density of states (\ref{eq:dens}),  
the function $F(t)$ is a so-called Basset function of first order~\cite{Basset} which must be analysed numerically.

\begin{figure}
	\includegraphics[width = 7cm, height = 5.6cm]{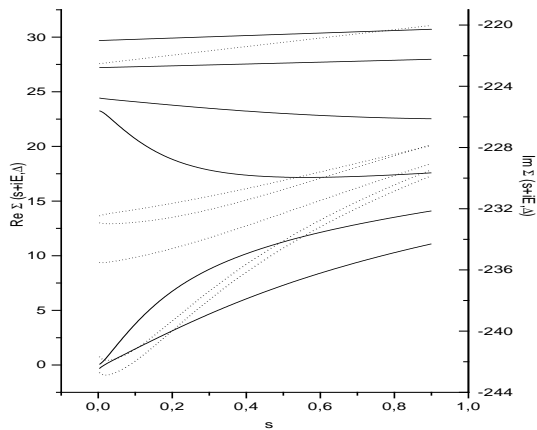}
	\caption{The complex transition coefficient $\Sigma (s+iE)$ plotted as a function of $s$ with fixed $E = 10.0$ for different values of the gap $\Delta$.
		The real components are drawn in solid curves and the imaginary parts in dotted ones. The corresponding gap values are in the order
                           from the topmost curve to the lowest curve: $\Delta = 0.0$, $\Delta = 3.0$, $\Delta = 4.5$, $\Delta = 4.9$, $\Delta = 5.1$ and $\Delta = 5.3$.
		For small values of $s$, the real components are nearly linear.}
	\label{fig:BassetSS}
\end{figure}

By performing the Laplace transform on the master equation (\ref{eq:master}), the right-hand side becomes $\Sigma (s) \sigma (s)$.
Assuming the density matrix \( \sigma \) to change slowly in a time scale of $\hslash / E$, the calculations can be restricted to the 
regime \( s \ll E \). In the normal-metal case, $\Sigma (s)$ varies only slowly as a function of small $s$, and therefore the zeroth-order 
approximation is reasonably good. However, when the energy $E/2$ is close to the gap energy $\Delta$, 
$\Sigma (s)$ has a strong $s$-dependence which is approximately linear for small $s$, as shown by  
Fig.~\ref{fig:BassetSS}. Using the linear approximation \( \Sigma (s) = a + b \cdot s \) and performing the inverse Laplace transform,
the right-hand side becomes \( \left( \Gamma + \Lambda \partial / \partial t\right) \sigma (t) \). Moving all the derivatives of the density matrix to the left-hand
side gives as a master equation
\begin{equation}
\label{eq:sigma3}
        (1- \Lambda) \frac{ \partial \sigma (t) }{ \partial t } + \frac{i}{ \hslash} \left[ \sigma (t), H_0 \right] = \Gamma \sigma (t),
\end{equation}
where \( \Gamma \) and \( \Lambda\) are tridiagonal matrices consisting of the zeroth- and first-order terms of the 
\( \Sigma (s) \) matrix, respectively. To guarantee the existence of the inverse of $(1-\Lambda)$, the elements of the $\Lambda$ coefficient 
are required to be small. This requirement is fulfilled when the tunnelling rate within the SET is small (that is, $|T^{\mathrm{L}}|,|T^{\mathrm{R}}| \ll 1$).
Multiplying equation (\ref{eq:sigma3}) from the left by \( (1-\Lambda)^{-1} \) the final form of the master equation is obtained as
\begin{equation}
        \frac{ \partial \sigma (t) }{ \partial t } = (1- \Lambda)^{-1} \left[ -\frac{i}{ \hslash} \left[ \sigma (t), H_0 \right] + \Gamma \sigma (t) \right] = \widetilde \Gamma \sigma (t).
\end{equation}
This describes a Markovian evolution of the density matrix and a sequential tunnelling approximation, which is a reasonable assumption
when the current in the SET is low.

The elements of the \( \Sigma (s)\) matrix are analysed by explicitly writing the Laplace transform
\begin{equation}
\label{eq:sigma}
        \Sigma (s) = \int_0^{\infty} \Sigma (t) e^{-st} dt = \int_0^{\infty} F(t)^2 e^{(-s+iE)t} dt.
\end{equation}
The linear approximation is done by calculating $\Sigma (s)$ for two values of $s=0.0$ and $s = 0.1$ and fitting a 
line. 

Elements of the matrix $\widetilde \Gamma$ describing the transitions within the qubit are proportional to a small mixing 
angle $\epsilon$ defined by \( \tan \epsilon = E_\mathrm{int} \sin \eta/(\Delta E + E_\mathrm{int} \cos \eta) \). 
By approximating these elements by zeroes, the matrix $\widetilde \Gamma$ separates into four parts:
one part describing the system when the qubit is in (diagonal) state $00$, one part when the qubit is in state $11$ and
two parts for the nondiagonal qubit states $01$ and $10$. The first two parts give raise to the measurement time as they 
describe the transistor's evolution for different qubit eigenstates. 
The coherence of the qubit is described by the strength of the non-diagonal elements, and thus the rate at which the non-diagonal
elements vanish gives the dephasing time. For the mixing time, one has to include the small mixing elements proportional to $\epsilon$.
To ease the treatment below, a Fourier transformation is performed in $m$-space: \( \sigma^k (t) = \sum_m e^{ikm} \sigma^m (t) \).

The measurement time is determined by the first two parts of the matrix $\widetilde \Gamma$. The equation corresponding
to the qubit state $00$ is
\begin{equation}
\label{eq:Meas0}
	\frac{d}{dt} \begin{pmatrix}
		\sigma_{00}^{N,k} \\
		\sigma_{00}^{N+1,k}
		\end{pmatrix}
=	\begin{pmatrix}
	-\Gamma_{L_{00}} & \Gamma_{R_{00}}  e^{ik} \\
	\Gamma_{L_{00}}  & - \Gamma_{R_{00}} 
	\end{pmatrix}
	\begin{pmatrix}
		\sigma_{00}^{N,k} \\
		\sigma_{00}^{N+1,k}
	\end{pmatrix},
\end{equation}
and the other part is a similar matrix equation for the qubit state $11$. These equations describe slowly damping (due to shot noise)
conductance peaks that propagate in time. The measurement has been performed once the peaks can be distinguished from each other,
and following the treatment in \cite{Schoen2000} the measurement time $t_\mathrm{ms}$ is defined as the time when the width of the peaks 
is smaller than the distance between their centers yielding
\begin{equation}
	t_{\mathrm{ms}} := \left( \frac{ \Gamma_{00} - \Gamma_{11}}{ \sqrt{2\Gamma_{00} f^0} + \sqrt{2\Gamma_{11} f^1}} \right)^2,
\end{equation}
where the group velocities are \( \Gamma_{i} := \Gamma_{L_{ii}} \Gamma_{R_{ii}} / ( \Gamma_{L_{ii}} + \Gamma_{R_{ii}}) \), with \( i = 0,1\).
The factors \( f^i = (\Gamma_{L_{ii}}^2 + \Gamma_{R_{ii}}^2 ) / (\Gamma_{L_{ii}}+\Gamma_{R_{ii}})^2\) are known as Fano factors \cite{Makhlin1999}.

The dephasing time is determined by the evolution of the nondiagonal qubit states given by the remaining
two parts of the matrix. Choosing $k=0$ has the effect of tracing out the $m$-variables 
and simplifying the corresponding submatrices
\begin{equation}
\label{eq:dephtime}
	\frac{d}{dt} \begin{pmatrix}
		\sigma_{01}^{N} \\
		\sigma_{01}^{N+1}
		\end{pmatrix}
=	\begin{pmatrix}
	i\Delta E^0 - \Gamma_{L_{01}} & \Gamma_{R_{01}} \\
	i\Delta E^1 + \Gamma_{L_{01}} & - \Gamma_{R_{01}} 
	\end{pmatrix}
	\begin{pmatrix}
		\sigma_{01}^{N} \\
		\sigma_{01}^{N+1}
	\end{pmatrix}.
\end{equation}
Again, these equations have plane wave solutions, and the corresponding eigenvalues
determine the dephasing time of the system. The time corresponding to each eigenvalue 
$\lambda$ is defined as $1/\mathrm{Re} (\lambda)$, and the dephasing time is chosen to be the longer of these
two times. The shorter time describes partial dephasing. In the superconducting case these 
must be evaluated numerically.

In order to analyse the mixing time of the system, also the mixing coefficients must be 
included in the $\widetilde \Gamma$ matrix. Once again, the SET degrees of freedom are
irrelevant and $k$ is chosen to be $0$. Now the matrix has eight eigenvalues: four of the eigenvalues describe
the dephasing, two describe the measurement and one eigenvalue is zero (describing the trace
preserving symmetry). The remaining real eigenvalue $\lambda_{\mathrm{mix}}$ gives the mixing time 
as the $t_{\mathrm{mix}} := 1/\lambda_{\mathrm{mix}}$, which must be analysed numerically.

If the mixing time is very large compared to the measurement time, the measuring device disturbs the
probability amplitudes of the qubit ($|a|^2$ and $|b|^2$ in $a|0\rangle + b|1\rangle$) only a little. The uncertainties of the charge $\Delta Q$ 
and its conjugate variable (phase or flux) $\Delta \Phi$ are linked to each other by the uncertainty principle
\( \Delta Q \cdot \Delta \Phi \geq \hslash/2 \). According to this principle, if the precision of the charge measurement is very high,
the phase becomes completely undetermined. The information that can be gathered from the SET contains no information 
of the phase and thus the precision of measuring the charge within the qubit ($|a|^2$ and $|b|^2$) can be very high. 
Furthermore, since charge is not disturbed by the measurement for $t_\mathrm{mix} \gg t_\mathrm{meas}$, we are essentially doing
the classic quantum non-demolition measurement.

The time scales are calculated for a specific set of parameters. With aluminium structures in mind, the superconducting gap 
of the SET is chosen to be $\Delta = 2.3 \, \mathrm{K}$, the capacitances $C_J = 8.0 \cdot 10^{-16} \, \mathrm{F}$, $C_T = 8.0 \cdot 10^{-17} \, \mathrm{F}$
and all the remaining capacitances are set to $1.0 \cdot 10^{-17} \, \mathrm{F}$. These fix the charging energies $E_{\mathrm{qb}} \approx 1.0 \, \mathrm{K} $, 
$E_{\mathrm{set}} \approx 5.2 \, \mathrm{K}$ and $E_{\mathrm{int}} \approx 0.13 \, \mathrm{K}$. 
The gate charges and tunnelling coefficients are chosen as $Q_\mathrm{qb} = 0.35$, $Q_\mathrm{set} = 0.15$, $\left| T^L \right|^2 = \left| T^R \right|^2 = 0.0025$.
To justify the charge-qubit approximation $E_{\mathrm{qb}} \gg E_{\mathrm{J}}$ the strength of the Josephson coupling is chosen as
$E_{\mathrm{J}} = 0.05 \, \mathrm{K}$. Finally, the biasing voltage is chosen as $eV = 16.5 \, \mathrm{K}$. This set of values gives the measurement 
time $t_{\mathrm{ms}} \approx 3.6 \cdot 10^{-10} \, \mathrm{s}$, the dephasing time $t_{\mathrm{\phi}} \approx 7.9 \cdot 10^{-10} \, \mathrm{s}$, and the mixing time $t_{\mathrm{mix}} \approx 2.3 \cdot 10^{-7} \, \mathrm{s}$.
The ratio between the mixing and measurement times is very high, i.e. $t_{\mathrm{mix}}/t_{\mathrm{ms}} \approx 625$. The same set of parameters for a normal-metal SET
($\Delta = 0.0 \, \mathrm{K}$) gives $t_{\mathrm{ms}} \approx  2.6 \cdot 10^{-6} \, \mathrm{s}$,  $t_{\phi} \approx  9.32 \cdot 10^{-10}\, \mathrm{s}$ and 
$t_{\mathrm{mix}} \approx  1.0 \cdot 10^{-7} \, \mathrm{s}$.

\begin{figure}[t]
        	\includegraphics[width=8cm, height = 7cm]{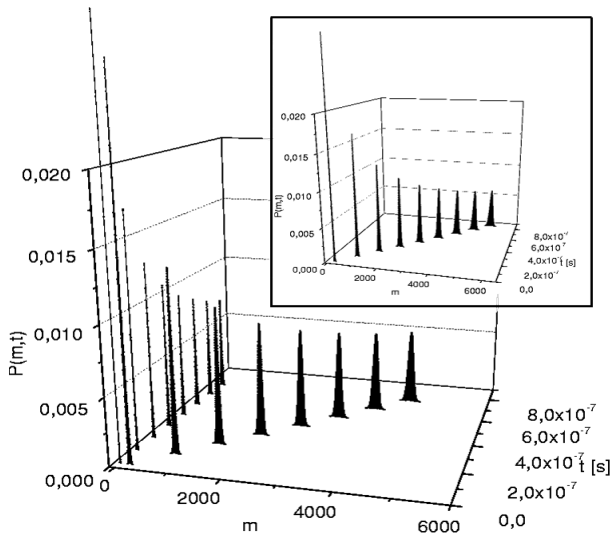}
	\caption{The probability $P(m,t)$ that $m$ electrons have tunnelled in a superconducting SET plotted as a function of time $t$
		for initial amplitudes $\sqrt{0.75}$ ($n=0$) and $\sqrt{0.25}$ ($n=1$). The parameters are for aluminium structures 
		and they are given in the text. The graph shows the separation of the two peaks, the faster corresponding 
		to the qubit's state $0$ and the slower to the state $1$. The two peaks are clearly distinguishable from each other, 
		and thus the measurement time is very small. The curve in the box shows the corresponding evolution for a normal-metal SET.}	
	\label{fig:Simul3}
\end{figure}

The system has been simulated numerically for the parameter values above, and the probabilities $P(m,t)$ for $m$ electrons having
been tunnelled during time $t$ are plotted for the superconducting and normal-metal cases in Fig.~\ref{fig:Simul3}. 
In the case of the superconducting SET, the measurement time is very small, as only one peak (corresponding to the qubit's state $n=0$) 
propagates in time and the other peak ($n=1$) is blocked by the presence of the gap. The mixing effects are small, as the peaks remain 
separated and the small visible spreading is caused mainly by the shot noise. For the normal-metal SET the measurement time is much longer than the mixing time and no 
separation of peaks is visible. Thus, for aluminium structures, the normal-metal SET is unlikely to be sensitive enough quantum measurement device in the high-voltage regime. 
Both for the normal-metal and superconducting SET the qubit state $|1\rangle$ or $|0\rangle$ determines the current through the SET
by shifting the energy levels in the middle island. However, due to the divergence in the superconducting density of states at the gap
this effect is strongly enhanced and leads to almost complete blocking of the current for $n=1$ in the superconducting case. In general,
this indicates that properties of the superconducting state can be very useful in performing sensitive quantum measurements.
In particular, the superconducting SET can be considered to constitute a very good quantum nondemolition measuring device as the mixing time is much larger than the measurement time 
$t_{\mathrm{mix}}/t_{\mathrm{ms}} \gg 1$. for parameter ranges accessible by aluminium structures. 

\emph{Acknowledgements} This work has been supported by the Academy of Finland (project 53903) under the Finnish
Centre of Excellence Project 2000-2005 (Project No. 44875, Nuclear and Condensed Matter Programme at JYFL).
We thank M. Aunola for useful discussions.

\end{document}